\documentclass[aps,prl,showpacs,twocolumn]{revtex4}
\usepackage{amssymb}
\usepackage{graphicx}
\usepackage{amsmath}
\usepackage{epsfig}
\usepackage{float}
\usepackage{color}
\usepackage{lscape}
\usepackage{pdflscape}
\usepackage[figuresright]{rotating}

\begin{document}

\title{Selective highly charged ions as the prospective optical clock candidates with quality factors larger than $10^{15}$}

\author{$^a$Yan-mei Yu \footnote[1]{E-mail: ymyu@aphy.iphy.ac.cn} and $^b$B. K. Sahoo \footnote[2]{E-mail: bijaya@prl.res.in}}
\affiliation{$^a$Beijing National Laboratory for Condensed Matter Physics, Institute of Physics, Chinese Academy of Sciences, Beijing 100190,China\\
$^b$Atomic, Molecular and Optical Physics Division, Physical Research Laboratory, Navrangpura, Ahmedabad 380009, India and\\
State Key Laboratory of Magnetic Resonance and Atomic and Molecular Physics, Wuhan Institute of Physics and Mathematics,
Chinese Academy of Sciences, Wuhan 430071, China}

\date{\today}

\begin{abstract}
We propose here a few selective highly charged ions (HCIs), namely Ni$^{12+}$ and Cu$^{13+}$, Pd$^{12+}$ and Ag$^{13+}$, that not only promise to be very high accurate optical clocks below $10^{-19}$ uncertainties, but also offer quality factors larger than $10^{15}$ and yet possess simple atomic structures for the experimental set-up. Moreover, these ions have strong optical magnetic-dipole (M1) transitions than the previously proposed HCI clocks. They can be used for the cooling and detection techniques. To demonstrate the projected fractional uncertainties below $10^{-19}$ level, we have estimated the typical orders of magnitudes due to many of the conventional systematics manifested in an atomic clock experiment, such as Zeeman, Stark, black-body radiation, and electric quadrupole shifts, by performing calculations of the relevant atomic properties.
\end{abstract}


\maketitle

Investigating properties of highly charged ions (HCIs) has a very interesting history of nearly one century long. Their spectra were used progressively for identifying abundant of elements in the solar corona and other astrophysical objects, describing nuclear reactions, etc. to great extent \cite{Raymond,Gillaspy,Prochaska-Nature-2003,Webb-PRL-2011}. However, only recently it has been recognized that HCIs are the potential
candidates for making ultra-high precision atomic clocks and probing many fundamental physics \cite{Yudin-PRL-2014,Derevianko-PRL-2012,Berengut-PRL-2012,Safronova-PRL-2014,dillip,Yu-PRA-2016}.
Relatively weaker forbidden transitions with narrow linewidths in HCIs are apt for excellent frequency standards as they are less sensitive to the external electromagnetic fields owing to extraordinarily contracted atomic orbitals compared to those of neutral or singly charged atoms. However, most of HCI transitions are in the x-ray region that are beyond the reach of the available narrow-bandwidth lasers. Only a very few HCIs, belonging to two specific categories, have been suggested for making optical clocks. One of the possibilities is the transition at the level crossings of two nearly degenerate states \cite{Berengut-PRL-2012,Derevianko-PRL-2012,Safronova-PRL-2014,dillip} and the other is the magnetic dipole (M1) transitions among the fine-structure and hyperfine manifolds \cite{Yudin-PRL-2014,Yu-PRA-2016}. Some of previously proposed HCIs have complicated atomic structures to carry out experiments and it is strenuous to perform high precision calculations in them for any theoretical studies, and also some suggested excited states of the clock transitions in these HCIs have short lifetimes which limit their quality factors ($Q$) to smaller values. Most of the earlier proposed HCI clock candidates lack appropriate optical cooling transitions. Laser trapping and cooling the clock candidates are immensely beneficial, which can enhance the interrogation time and reduce the Doppler shifts sharply for the ultra-high precision measurements. Therefore, it is desirable to find out HCIs that can provide both the clock and cooling transitions in the optical region.

\begin{figure}[t]
\begin{center}
  \includegraphics[width=8cm]{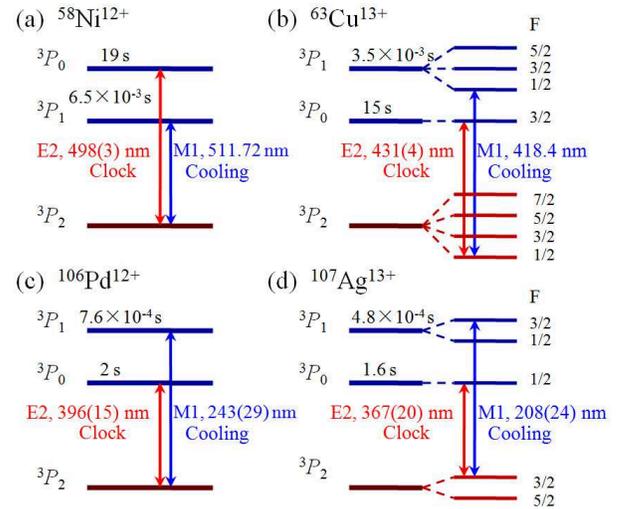}
\end{center}\vspace{-5mm}
\caption{Schematic diagrams of the low-lying atomic and hyperfine energy levels of the (a) $^{58}$Ni$^{12+}$, (b) $^{63}$Cu$^{13+}$, (c) $^{106}$Pd$^{12+}$, and (d) $^{107}$Ag$^{13+}$ HCIs. Intercombination lines that are apt for the clock and cooling transitions are shown in the red and blue colors, respectively. The estimated wavelengths and lifetimes of the excited states in the clock and cooling transitions are mentioned.}
\label{fig1}
\end{figure}

In this Letter, we propose a new class of HCIs for optical atomic clocks aiming at the 10$^{-19}$ level of accuracy. Particularly, the S-like Ni$^{12+}$ and Cu$^{13+}$ and Se-like Pd$^{12+}$ and Ag$^{13+}$ HCIs having the $3p^4$ and $4p^4$ ground-state configurations, respectively, appear to hold unique features to be suitable as the optical frequency standards. We mention some of the important features of these new clock candidates here: (i) The E2 transitions between the $^3P_2-^3P_0$ intercombination levels of the ground states of these ions, as shown in Fig. \ref{fig1}, are found to be in the optical range with $Q$ values ranging $10^{15}-10^{16}$. On the other hand, these transitions are not so highly forbidden like the electric octupole (E3) clock transition of the $^{171}$Yb$^+$ ion \cite{Huntemann}, so they can be directly interrogated within a stipulated time. (ii) The intercombination lines offer several strong M1 transitions in the optical region that can be used for the ion cooling, state manipulation and states detection techniques. (iii) We demonstrate later that most of the typical systematics of an atomic clock experiment such as black-body radiation (BBR), Zeeman, Stark, quadrupole shifts, etc. are estimated to be negligibly small in these ions projecting their maximum fractional uncertainties to their clock frequencies ($\nu$) below $10^{-19}$ level. (iv) The atomic level structures of the undertaken ions are comparatively simpler than the earlier proposed HCIs for the optical clocks. Two isotopes have nuclear spin $I=0$, which is advantageous to suppress the second-order Zeeman effect. (v) The last but not the least to mention here that the Ni and Cu HCIs are abundant in the astrophysical objects \cite{Prochaska-Nature-2003}; thus one can make a comparative analysis between the astrophysical observations with the clock measurements to infer about temporal variation of the fine structure constant \cite{Webb-PRL-2011}.

The basic spectroscopy of the aforementioned S-like HCIs have been investigated before for their extensive usages in the identification and interpretation of the emission lines in the astrophysics, fusion reactors, plasma diagnostics, etc. \cite{Sugar-JPCRD-1990,Shirai-JPCRD-1991,Chou-ADNDT-1996,Bhatia-ADNDT-1998,Trabert-JPB-2009}. However, most of these studies are concentrated on the allowed transitions, since their optical signals can be observed easily. The forbidden transitions between the $^3P_0$ and $^3P_2$ states, which are also important from the astrophysics point of view, have received less attentions. On the other hand, the spectroscopy of the considered Se-like ions are not well known. We have performed calculations of the excitation energies (EEs) and
lifetimes of the excited states using the multi-configuration Dirac-Hartree-Fock (MCDF) method of the GRASP2K program \cite{GRASP}, that are useful for the clock and cooling transitions of the above ions, and give them in Table \ref{tab:energy}. The energies are compared with the experimental values quoted in the National Institute of Science and Technology (NIST) database \cite{NIST}. We have considered the Dirac-Coulomb (DC) interaction Hamiltonian along with the Breit and lower-order QED effects in these calculations. The calculated energies in Ni$^{12+}$ and Cu$^{13+}$ are in good agreement with the experimental values \cite{NIST}, from which the wavelengths of the proposed $^3P_0- ^3P_2$ clock transition are estimated to be about 2\% accuracy. On this basis, we assume that our calculations for the Pd$^{12+}$ and Ag$^{13+}$ HCIs are of similar accuracies. We have also determined EEs of these ions by employing the Fock-space based relativistic coupled-cluster (FSCC) method of the DIRAC package \cite{DIRAC} and find both the calculations agree to each other reasonably well (see the Supplemental material). From the above table, the lifetimes of the $^3P_2$ states of Ni$^{12+}$ and Cu$^{13+}$ are found to be more than ten seconds, while these states have about a few seconds lifetimes in Pd$^{12+}$ and Ag$^{13+}$. From this analysis, we have chosen the $^3P_2-^3P_0$ transition in $^{58}$Ni$^{12+}$ and $^{106}$Pd$^{12+}$ that have $I=0$, the $|^3P_2,~F=1/2\rangle-|^3P_0,~F=3/2\rangle$ transition in $^{63}$Cu$^{13+}$ with $I=3/2$, and the $|^3P_2,~F=3/2\rangle-|^3P_0,~F=1/2\rangle$ transition in $^{107}$Ag$^{13+}$ with $I=1/2$ as the possible optical clock transitions. In the above table, we also give EEs and lifetimes of other excited states with the $^3P_1$, $^1D_2$, and $^1S_0$ configurations. The $^3P_1-^3P_2$ M1 transitions in the Cu$^{13+}$, Pd$^{12+}$, and Ag$^{13+}$ ions are also in the optical region. Therefore, they can be used for the cooling and state manipulation techniques.

\begin{table}[t]
\caption{The estimated values of EEs, lifetimes ($\tau$), wavelengths ($\lambda$) and transition rates ($A^O_{ki}$) due to the forbidden channels ($O=$M1 or E2) of the excited states to the ground states of the considered HCIs. Energies are compared with the available NIST data \cite{NIST}. Other contributing transition rates to the lifetimes can be found in the Supplemental material. Uncertainties are quoted within the parentheses.  Numbers in the square brackets represent powers of 10. \label{tab:energy}}
{\setlength{\tabcolsep}{2.0pt}
\begin{tabular}{ lllll  c c } \hline\hline
 Level	      & \multicolumn{2}{c}{EE (in cm$^{-1}$)} & $\lambda$ & $O$ & $A^O_{ki}$ & $\tau$ \\
              &  This work & NIST \cite{NIST} & nm & & s$^{-1}$ & s \\ \hline
 \multicolumn{7}{c}{Ni$^{12+}$ ion ($3p^4$ configuration)} \\
           $^3P_1$ &19560(20)   &19541.8      & 511.72 &M1 & 154 & $6.5[-3]$ \\
           $^3P_0$ &20251(450)  &20060(100)   & 498(3) &E2 & 0.03 & 19 \\
           $^1D_2$ &47512(2120) &47032.9      & 212.61 &M1 & 261  & $3.6[-3]$     \\
           $^1S_0$ &99377(5130) &97836.2      & 102.21 &E2 & 3.37 & $3.8[-4]$       \\
           \multicolumn{7}{c}{Cu$^{13+}$ ion ($3p^4$ configuration)} \\
           $^3P_0$ &23457(350)  &23192(100)   & 431(4) &E2 & 0.06 & 15     \\
           $^3P_1$ &23922(260)  &23897        & 418.4  &M1 & 283  & $3.5[-3]$            \\
           $^1D_2$ &53014(1560) &52540        & 190.3  &M1 & 421  & $2.2[-3]$     \\
           $^1S_0$ &109356(4020)&107902       & 92.7   &E2 & 4.34 & $2.4[-4]$      \\
            \multicolumn{7}{c}{Pd$^{12+}$ ion ($4p^4$ configuration)} \\
           $^3P_0$ &25275(470)  &             & 396(15) &E2 & 0.5  &  2    \\
           $^3P_1$ &41190(2470) &             & 243(29) &M1 & 1260 & $7.6[-4]$   \\
           $^1D_2$ &60037(500)  &             & 167(3)  &M1 & 1155 & $8.5[-4]$     \\
           $^1S_0$ &119562(2390)&             & 83.6(3) &E2 & 17.1 & $1.2[-4$        \\
           \multicolumn{7}{c}{Ag$^{13+}$ ion ($4p^4$ configuration)} \\
           $^3P_0$ &27242(760)  &             & 367(20) &E2 & 0.6  & 1.6      \\
           $^3P_1$ &48028(2760) &             & 208(24) &M1 & 1977 & $4.8[-4]$      \\
           $^1D_2$ &67514(540)  &             & 148(2)  &M1 & 1736 & $5.6[-4]$      \\
           $^1S_0$ &134012(2540)&             & 74.6(3) &E2 & 21.9 & $8.2[-5]$       \\ \hline\hline
\end{tabular}}
\end{table}

\begin{table}[]
\caption{Values of $\nu$, $Q$, $\Theta$ (in a.u.), $\alpha^{E1}$ (in a.u.), $\alpha^{M1}$ (in a.u.), $A_{hf}$ (in MHz) and $B_{hf}$ (in MHz) are listed along with the fractional differential (denoted by $\delta$ symbol) systematic shifts for the respective HCIs. We have used nuclear magnetic dipole moment ($\mu_I$) and electric quadrupole moment ($Q_I$) for $^{63}$Cu as $\mu_I=2.223\mu_N$ and $Q_I=-0.211b$, and for $^{107}$Ag as $\mu_I=-1.135\mu_N$ and $Q_I=0.98b$ in the determination of the $A_{hf}$ and $B_{hf}$ values. Here, $\alpha^{M1}$ values are given for the $^a$$J=2$, $^b$$F=1/2$, and $^c$$F=3/2$ states. Numbers in the square brackets represent powers of 10. \label{tab:clockS}}
{\setlength{\tabcolsep}{2pt}
\begin{tabular}{lcccc }\hline\hline
Items                     &$^{58}$Ni$^{12+}$ &$^{106}$Pd$^{12+}$ &$^{63}$Cu$^{13+}$ &$^{107}$Ag$^{13+}$ \\ \hline
$I$                       & 0                &0                  &3/2               &1/2                \\
$v$                       &$6.01[14]$        &$7.64[14]$         &$6.95[14]$        & $8.06[14]$        \\
$Q$                       &1.1[16]           &1.5[15]            &1.0[16]           &1.3[15]            \\
$A_{hf}$($^3P_2$)         &                  &                   &10564             &$-1594$           \\
$B_{hf}$($^3P_2$)         &                  &                   &-204              &0                 \\
$\Theta(^ 3P_2)$          &0.1465            &0.1670             &0.1260            &0.1382             \\
$\alpha^{E1}$($^3P_2$)    &0.2736            &0.6531             &0.2295            &0.5617             \\
$\alpha^{E1}$($^3P_0$)    &0.2751            &0.6548             &0.2306            &0.5625             \\
$\alpha^{M1}$($^3P_2$)    &2.10$^a$          &1.35$^a$           &-4.62[5]$^b$      &$-3.71[5]$$^c$     \\
$\alpha^{M1}$($^3P_0$)    &$-149.85$         &4.66               &103.76            &$-3.58$           \\
$\delta E^{(2)}_{Zeem}/v$ &$3.77[-23]$       &$-6.60[-25]$       &$9.89[-20]$       &$-6.85[-20]$      \\
$\delta E_{Stark}/v$      &$-1.52[-24]$      &$-1.43[-24]$       &$-9.93[-25]$      &$-7.33[-25]$      \\
$\delta E^{E1}_{BBR}/v$   &$-2.10[-20]$      &$-1.67[-20]$       &$-1.37[-20]$      &$-1.01[-20]$      \\
$\delta E^{M1}_{BBR}/v$   &$-2.50[-20]$      &$-3.85[-21]$       &$7.88[-20]$       &$1.60[-20]$       \\ \hline\hline
\end{tabular}}
\end{table}

It has also been one of the most demanding tasks so far to find out suitable cooling transitions for the HCI clocks. We have come up with a few step-wise strategy of cooling mechanism to bring down the proposed HCIs to the desired low temperature for the clock frequency measurements. As known HCIs can be produced using one of the procedures like accelerator, high-power laser, electron cyclotron resonance, electron beam ion sources etc. techniques. Initially these ions can be trapped at very high temperature, in the order of megakelvin, then they can be decelerated in the traditional approach to reach temperature around $10^5$K. At this stage, one may adopt the evaporation technique to bring the temperature down to $10^4$K with the storage time up to 1000s \cite{Hobein-PRL-2011}. Following this, sympathetic cooling technique can be to applied to decrease the temperature down to milliKelvin as has been demonstrated in Ref.
\cite{Schmoger-Science-2015}. In this process another species is trapped simultaneously with the ion matching with its mass to charge ratio. In our case, the ratios for the Pd$^{12+}$ and Ag$^{13+}$ HCIs match well with the Be$^{+}$ ion. Subsequently, the narrow $^3P_1-^3P_2$ lines in Cu$^{13+}$, Pd$^{12+}$, and Ag$^{13+}$, as shown in Fig. \ref{fig1}, can be used for the laser cooling to achieve temperature about nK. This can suppress the uncertainties due to the second-order Doppler effects in the proposed HCI clocks.

We analyze the most commonly transpired systematics in the atomic clock experiments for the considered HCIs. We list many properties of the atomic states associated with the clock transitions including $I$ values of the relevant isotopes in Table \ref{tab:clockS} and give typical orders of magnitudes of the systematics deducing from them. It to be noted that these are the absolute values, but more precise values of the clock frequencies can be obtained by determining their uncertainties when the actual measurements are carried out. The $Q$ values are estimated from the calculated lifetimes of the excited states given in Table \ref{tab:energy}. We have evaluated the electric dipole polarizabilities ($\alpha^{E1}$) and quadrupole moments ($\Theta$) adopting the finite-field approach in the FSCC method \cite{DIRAC} and magnetic dipole ($A_{hf}$) and electric quadrupole ($B_{hf}$) hyperfine structure constants using the MCDF method \cite{GRASP}. More details of these results can be found in the Supplemental materials. We also give magnetic dipole polarizabilities ($\alpha^{M1}$), that may contribute significantly to the proposed clock transitions.

The first-order Zeeman shift is defined by $\delta E^{(1)}_{Zeem}=\delta g \delta M \mu_B B$ for the external magnetic field $B$, Bohr magneton $\mu_B$ and differential value of Lande $g$-factors $\delta g$ and $M$-quantum number of the states. The $\delta E^{(1)}_{Zeem}$ shift can be eliminated by choosing the $M_J=0\rightarrow M_J=0$ transition in Ni$^{12+}$ and Pd$^{12+}$ or measuring for all the $M$-components in Cu$^{13+}$ and Ag$^{13+}$. However, the second-order shift, $\delta E^{(2)}_{Zeem}=-\frac{1}{2} \delta\alpha^{M1}B^2$, can be significant, wherein $\delta\alpha^{M1}$ is the differential value of $\alpha^{M1}$ between the clock states. By defining the M1 operator $O^{M1} = (\vec L + 2 \vec S) \mu_B$ with $L$ and $S$ as the orbital and spin angular momentum operators respectively, we evaluate $\alpha^{M1}$ for the $|\gamma J M_J\rangle$ state as
\begin{eqnarray}\label{eq:alphaM1J}
\alpha^{M1}(J)= - \frac{2}{3(2J+1)} \sum_{J'} \frac{ |\langle J||O^{M1}||J'\rangle|^2}{E_J-E_{J'}},
\end{eqnarray}
with the energies $E$ and the reduced matrix element
\begin{eqnarray}\label{eq:alphazJ}
\langle J||O^{M1}||J'\rangle &=& \mu_B \sqrt{ S(S+1)(2S+1) (2J+1)} \nonumber \\ & \times & \sqrt{(2J'+1)} \Bigg\{ \begin{array}{c c c} L&S&J\\1&J'&S  \end{array} \Bigg\} .
\end{eqnarray}
For the hyperfine level $|F,M_F\rangle$, we estimate $\alpha^{M1}$ by considering only the dominant contributions from the hyperfine manifolds of the same principal ($n$) and angular momentum ($J$) states and neglecting contributions from other angular momentum states as
\begin{eqnarray}\label{eq:alphaM1F}
\alpha^{M1}(nIJF)= - \frac{2}{3(2F+1)} \sum_{F'} \frac{ |\langle nIJF||O^{M1}||nIJF'\rangle|^2}{E_{nIJF}-E_{nIJF'}},
\end{eqnarray}
in which the reduced matrix element of $O^{M1}$ is given by
\begin{eqnarray}\label{eq:alphazF}
\langle nIJF||O^{M1}||nIJF'\rangle &=& \mu_B \sqrt{ J(J+1)(2J+1)(2F+1)} \nonumber \\ & \times & \sqrt{(2F'+1)} \Bigg\{ \begin{array}{c c c} I & J &F \\1&F'& J  \end{array} \Bigg\} g_J , \ \  \
\end{eqnarray}
in which $g_J$ is evaluated by
\begin{eqnarray}\label{eq:gJ}
g_J=1+\frac{J(J+1)-L(L+1)+S(S+1)}{2J(J+1)}.
\end{eqnarray}
We have determined hyperfine energy levels as
\begin{eqnarray}\label{eq:EnJF}
E_{nJF}&=&\frac{1}{2} A_{hf} K  \nonumber \\
&+&B_{hf} \frac{(3/2)K(K+1)-2I(I+1)J(J+1)}{4I(2I-1)J(2J-1)},
\end{eqnarray}
where $K$=$F(F+1)-I(I+1)-J(J+1)$. Assuming a typical value of $B$=$5\times10^{-8}$T \cite{Derevianko-PRL-2012}, $\delta E^{(2)}_{Zeem}$ is found to be below $10^{-19}$. However, we find that the second-order Zeeman shift in the $^{63}$Cu$^{13+}$ and $^{107}$Ag$^{13+}$ HCIs are slightly larger due to small energy difference between the hyperfine levels. To suppress this systematics in these two ions, it may be needed to calibrate the magnetic field more stringently.

The $\alpha^{E1}$ value of a HCI scales as $1/Z_i^4$, where $Z_i$ is the residual nuclear charge. Our calculation gives differential scalar polarizability values between the states associated with the clock transitions of Ni$^{12+}$, Cu$^{13+}$, Pd$^{12+}$, and Ag$^{13+}$ as $\delta \alpha_0^{E1} \approx10^{-3}$ atomic unit (a.u.). This tiny $\delta \alpha_0^{E1}$ values lead to negligibly small fractional differential Stark shifts, $\delta E_{Stark}$=$-\delta \alpha_0^{E1} {\cal E}^2/2$, to the clock transitions with the typical electric field strength ${\cal E}$=$10 \ V/m$. Also, the light shifts that could be caused by the probe and cooling lasers can be strongly suppressed owing to very small $\delta \alpha_0^{E1}$ values. Here we have assumed that contribution due to the tensor components, which are one order smaller than their scalar polarizability values in the states with $J$=2, can be either averaged out by performing measurements in all the M-components or suppressed by using slightly weaker probe light intensity \cite{Dube-PRA-2013}.

The differential BBR shift at the room temperature, $T$=300K, due to the E1 channel for the clock transitions are estimated using the expression
\begin{equation}\label{eq:BBRE1}
\delta E^{E1}_{BBR}=-\frac{1}{2}(831.9V/m)^2 \left [\frac{T(K)}{300} \right]^4\delta \alpha_0^{E1}
\end{equation}
and found to be much lower than $10^{-19}$ level. Similarly, the BBR shift due to other dominant M1 channel of the $|\eta \rangle$ state, say, can be estimated using the formula
\begin{eqnarray}
\Delta E^{M1}_{BBR}&=& -\frac{\mu_0(K_BT)^2}{\pi^2(c\hbar)^3}  \sum_{\beta} \bigg[ |\langle \eta|| O^{M1}|| \beta \rangle|^2 \omega_{\eta \beta} \nonumber \\ & \times & \int^{\infty}_0 d\omega \frac{\omega^3}{(\omega_{\eta \beta}^2-\omega^2)(\exp^{\hbar\omega/K_BT}-1)}\bigg] ,
\end{eqnarray}
where $\mu_0$, $K_B$, $\hbar$, and $c$ are the magnetic permeability, Boltzmann's constant, Plank's constant, and speed of light, respectively. The respective expressions of $\langle \eta|| O^{M1}|| \beta \rangle$ for the atomic and hyperfine levels are used from Eqs. (\ref{eq:alphazJ}) and (\ref{eq:alphazF}) to estimate the M1 BBR shifts of the states involved with the clock transitions. Then, the BBR shift of the clock transition is determined as
\begin{equation}\label{eq:BBRE1}
\delta E^{M1}_{BBR}=\Delta E^{M1}_{BBR}(^3P_0)-\Delta E^{M1}_{BBR}(^3P_2),
\end{equation}
which are found to be below 10$^{-19}$ at the room temperature in all the four HCIs.

The electric quadrupole shift, $\Delta E_{\Theta}=-\Theta \mathcal E_{zz}/2$ caused by the gradient of the applied electric field $\mathcal E_{zz}$ in the $z$-direction on a state with quadrupole moment $\Theta$, could be a major systematic in the atomic clocks. Our choice of upper state as $^3P_0$ and lower state as $^3P_2$ ($F=1/2$) in Cu$^{13+}$ warrants $\Delta E_{\Theta}=0$ since $\Theta=0$ in these states. However, $\Delta E_{\Theta}$ values in the lower clock states in Ni$^{12+}$, Pd$^{12+}$, and Ag$^{13+}$ are substantially large, about $10^{-15}$ with respective to $\nu$ for the typical magnitude of $\partial E_z/\partial z \approx10^8$V/m$^2$ due to nonzero values of $\Theta$. It is possible to nullify $\Delta E_{\Theta}$ in these states by performing measurements in all their $M$-components and averaging out as has been demonstrated in \cite{Dube-PRA-2013}.

From the wavelengths and lifetimes of the excited states listed in Table \ref{tab:energy} and magnitudes of all the major systematics mentioned in Table \ref{tab:clockS}, it is evident that the selected S-like Ni$^{12+}$ and Cu$^{13+}$ and the Se-like Pd$^{12+}$ and Ag$^{13+}$ ions are the potential HCIs for the optical for the optical atomic clocks with $Q$-factors larger than $10^{15}$ and fractional systematics below $10^{-19}$. These HCIs also have several exceptional advantages as clocks, such as availability of strong transitions in optical range facilitating to adopt the laser cooling, reliable initial state preparation and efficient state detection mechanisms, possibility to infer variation of the fine structure constant comparing with astrophysical observations etc., apart from having simpler atomic energy levels. Most importantly these ions can be easily produced using the presently available facilities.

Y.Y. was supported by the CAS XDB21030300, the NKRD Program of China (2016YFA0302104), and the National Natural Science Foundation of China under Grant No. 91536106. B.K.S. acknowledges financial support from Chinese Academy of Science (CAS) through the PIFI fellowship under the project number 2017VMB0023 and partly by the TDP project of Physical Research Laboratory (PRL), Ahmedabad.

\newpage

\begin{thebibliography}{}

\bibitem{Raymond}
J. C. Raymond, {\it Highly Charged Ions in Astrophysics}, R. Marrus (eds), Physics of Highly-Ionized Atoms, NATO ASI Series (Series B: Physics), vol 201. Springer, Boston, MA (1989).

\bibitem{Gillaspy}
J. D. Gillaspy, J. Phys. B: At. Mol. Opt. Phys. {\bf 34}, R93 (2001).

\bibitem{Prochaska-Nature-2003}
J. X. Prochaska, J. Christopher Howk, and A. M. Wolfe, Nature {\bf 423}, 57 (2003).

\bibitem{Webb-PRL-2011}
J. K. Webb, J. A. King, M. T. Murphy, V.V. Flambaum, R. F. Carswell, and M. B. Bainbridge, Phys. Rev. Lett. {\bf 107}, 191101 (2011).

\bibitem{Yudin-PRL-2014}
V. I. Yudin, A. V. Taichenachev, and A. Derevianko, Phys. Rev. Lett. {\bf 113}, 233003 (2014).

\bibitem{Derevianko-PRL-2012}
A. Derevianko, V. A. Dzuba, and V.V. Flambaum, Phys. Rev. Lett. {\bf 109}, 180801 (2012).

\bibitem{Berengut-PRL-2012}
J. C. Berengut, V. A. Dzuba, V.V. Flambaum, and A. Ong, Phys. Rev. Lett. {\bf 109}, 070802 (2012).

\bibitem{Safronova-PRL-2014}
M. S. Safronova, V. A. Dzuba, V. V. Flambaum, U. I. Safronova, S. G. Porsev, and M. G. Kozlov, Phys. Rev. Lett. {\bf 113}, 030801 (2014).

\bibitem{dillip}
D. K. Nandy and B. K. Sahoo, Phys. Rev. A {\bf 94}, 032504 (2016).

\bibitem{Yu-PRA-2016}
Y.M. Yu and B. K. Sahoo, Phys. Rev. A {\bf 94}, 062502 (2016).

\bibitem{Huntemann}
N. Huntemann, C. Sanner, B. Lipphardt, Chr. Tamm, and E. Peik, Phys. Rev. Lett. {\bf 116}, 063001 (2016).

%
%
\bibitem{Sugar-JPCRD-1990}
J. Sugar and A. Musgrove, J. of Phys. and Chem. Ref. Data {\bf 19}, 527 (1990).

\bibitem{Shirai-JPCRD-1991}
T. Shirai, T. Nakagaki, Y. Nakai, J. Sugar, K. Ishii, and K. Mori, J. of Phys. and Chem. Ref. Data {\bf 20}, 1 (1991).

\bibitem{Chou-ADNDT-1996}
H. S. Chou, J.-Y. Chang, Y.-H. Chang, and K.-N. Huang, At. Data Nucl. Data Tables {\bf 62}, 77 (1996).

\bibitem{Bhatia-ADNDT-1998}
A. K. Bhatia and G.A. Doschek, At. Data Nucl. Data Tables {\bf 68}, 49 (1998).


\bibitem{Trabert-JPB-2009}
E. Tr\"{a}bert, J. Hoffmann, C. Krantz, A. Wolf, Y. Ishikawa and J. A. Santana, J. Phys. B: At. Mol. Opt. Phys. {\bf 42}, 025002 (2009).

\bibitem{GRASP}
P. J\"{o}nsson, G. Gaigalas, J. Biero, C. F. Fischer, and I. P. Grant, Comput. Phys. Comm. {\bf 177}, 597 (2007).

\bibitem{DIRAC}
{DIRAC}, a relativistic ab initio electronic structure program, Release {DIRAC13} (2013), written by L. Visscher, H. J. Aa. Jensen, R. Bast, T.
Saue, and Collaborators (http://www.diracprogram.org).

\bibitem{NIST}
http://physics.nist.gov/PhysRefData/ASD/levels\_form.html.

\bibitem{Hobein-PRL-2011}
M. Hobein, A. Solders, M. Suhonen, Y. Liu, and R. Schuch, Phys. Rev. Lett. {\bf 106}, 013002 (2011).

\bibitem{Schmoger-Science-2015}
L. Schm\"{o}ger, O. O. Versolato, M. Schwarz, et al., Science {\bf 347}, 1233 (2015).



\bibitem{Dube-PRA-2013}
P. Dub\'{e}, A. A. Madej, Z. Zhou, and J. E. Bernard, Phys. Rev. A {\bf 87}, 023806 (2013).

\end{thebibliography}
\end{document}